\documentstyle[12pt]{article}
\advance \textheight by \topskip
\makeatletter
\def\eqnarray{\stepcounter{equation}\let\@currentlabel=\theequation
\global\@eqnswtrue
\global\@eqcnt\z@\tabskip\@centering\let\\=\@eqncr
$$\halign to \displaywidth\bgroup\@eqnsel\hskip\@centering
  $\displaystyle\tabskip\z@{##}$&\global\@eqcnt\@ne
  \hfil$\displaystyle{\hbox{}##\hbox{}}$\hfil
  &\global\@eqcnt\tw@ $\displaystyle\tabskip\z@
  {##}$\hfil\tabskip\@centering&\llap{##}\tabskip\z@\cr}
\def\tdot#1{\raise-0.1ex\hbox{$\mathop{#1}\limits^{...}$}}

\begin{document}

\begin{titlepage}
\hbox to \hsize{\hfil hep-th/9404032}
\hbox to \hsize{\hfil IHEP 92--87}
\hbox to \hsize{\hfil June, 1992}
\vfill
\large \bf
\begin{center}
B\"ACKLUND TRANSFORMATION \\
FOR INTEGRABLE SYSTEMS
\end{center}
\vskip 1cm
\normalsize
\begin{center}
{\bf A. N. Leznov\footnote{E--mail: leznov@mx.ihep.su}}\\
{\small Institute for High Energy Physics, 142284 Protvino, Moscow Region,
Russia}
\end{center}
\vskip 2.cm
\begin{abstract}
\noindent
We establish an explicit form of the Backlund transformation for
the most known integrable systems.
\end{abstract}
\vfill
\end{titlepage}

1. In this paper the Backlund transformation and its integration for the most
known and applicable integrable systems is obtained. Here, by the
Backlund transformation we mean any nonlinear mapping which
transfers any given solution into another one. However, in this we do
not investigate the properties of the transformation, its
geometrical interpretation (if any), etc. To prove its validity one can make a
direct check which use only one operation -- differentiation.

As a hint for obtaining the Backlund transformation the author used a
purely algebraic method for construction of the soliton type solutions,
see \cite{1,2}, modified for the case of the solvable algebras, \cite{3}.

2.The starting point of construction below use the following two facts.
The integrable systems under consideration admit the transformation $s$:

$$ \theta \Rightarrow \tilde \theta \equiv S \theta =
F(\theta, \theta^{(1)}_i...\theta^N_i), \qquad S^N\not=1.
$$
Here ${\theta}$ and ${\tilde \theta}$ are unknown functions (variables)
satisfying the corresponding PDEs, $\theta^\nu_i \equiv \frac{\partial^{\nu}
\theta}{\partial x^{\nu}_i}.$

There is the obvious solution of the nonlinear system in question which
depends on a set of arbitrary functions.

The soliton type solutions, reductions related with the discrete groups,
solutions with definite boundary conditions are defined by the special
choice of arbitrary functions mentioned above.
Let us note that $\theta_0$ is a solution of a linear system of
partial differential equations and it can be presented as a parametric
integral on the plane of the complex variable $\lambda$.
This circumstance is just the main reason for applying to the integrable
systems the methods of the theory of functions of complex variables, the
technics of the Riemann problem, and, at last, the methods of the inverse
scattering problem.

  3. Here we give a list of integrable systems together with the Backlund
transformations for them and the corresponding solutions.

1. Hirota equation
\begin{eqnarray}
v'+\alpha(\tdot v-6uv \dot v)-i\beta(\ddot v-2v^2u)+\gamma
\dot v+ i\delta v & = & 0, \nonumber\\
u'+\alpha (\tdot u-6uv\dot u)+i\beta (\ddot u-2u^2v)
+ \gamma \dot u- i\delta u  & = & 0;
\end{eqnarray}
\[
'\equiv \frac{\partial}{\partial t}, \qquad
. \equiv \frac{\partial}{\partial x};
\]
\[
\tilde v \equiv sv = \frac{1}{u}, \qquad \tilde u \equiv su =
u(uv-\ddot{\mbox{ln}u}),\qquad v_0=0,
\]
\[
u'_0+\alpha \tdot u_0+i\beta \ddot u_0 + \gamma \dot u_0 - i\delta u_0=0.
\]

In this and in the other cases the main role will be played by the principal
minors of the following matrix:
\[
\left(
\begin{array}{cccc}
\phi^s & \phi^{s+1} & \phi^{s+2} & \ldots \\
\phi^{s+1} & \phi^{s+2} & \phi^{s+3} & \ldots \\
\phi^{s+2} & \phi^{s+3} & \phi^{s+4} & \ldots \\
\ldots & \ldots & \ldots & \ldots
\end{array}
\right)
\]
To denote the principal minors of these matrices we shall use the symbol
$ D_r^n $. Here $n$ is the rank of the matrix and $r$ is the symbol of its
element of left upper corner. For the solution of the Backlund transformation
we have
\begin{equation}
v_n=(-1)^n \frac{D_0^{n-1}}{D_0^n}\, , \qquad
u_n=(-1)^{n+1} \frac{D_0^{n+1}}{D_0^{n+1}}
\end{equation}
The methods of theory function of complex variables gives the same
expression where the role $ D_0^n $ play the nonlocal integral
\begin{equation}
D_0^n=\int d\lambda_1...d\lambda_n c(\lambda_1)...c(\lambda_n)
W_n^2(\lambda_1,...,\lambda_n),
\end{equation}
where $W_n(\lambda)$ is the Vandermonde determinant; and $c(\lambda)$
is the integrand in the representation for $u_0$.

2. Nonlinear Schr\"odinger equation

a)
\begin{eqnarray}
q'+ \ddot q - 2rq^2 = 0   & \tilde q = \frac{1}{r} &
\tilde r = r[rq-\ddot{\mbox{ln} r}] ;\nonumber\\
-r'+ \ddot r -2qr^2 = 0  & q_o = 0 &  r'_0= \ddot r_0.\label{4}
\end{eqnarray}
The solution of the Backlund transformation is the same as in the previous
section.

b)
\begin{eqnarray}
q'+\ddot q + 2(rq)\dot q = 0 & \tilde q = \frac{1}{r} &
\tilde r = r[(rq) + \dot {\mbox{ln}\frac{r}{\dot r}}];\nonumber\\
-r'+ \ddot r - 2(rq) \dot r = 0 & q_0=0 & r'_0 = \ddot r_0.
\end{eqnarray}
The solution of the Backlund transformation is as follows
\begin{equation}
q_n=(-1)^n\frac{D_1^{n-1}}{D_0^n}\, , \qquad
r_n=(-1)^{n+1}\frac{D_0^{n+1}}{D_1^{n+1}}
\end{equation}
c)
\begin{eqnarray}
q' + \ddot q - 2\dot (rq^2) = 0 & \tilde q = r, &
\tilde r = q -\dot(\frac {1}{r}) ;\nonumber\\
-r' + \ddot r + 2\dot(r^2q)=0    & q_0 = 0  & r'_0 = \ddot r_0.\label{7}
\end{eqnarray}
The solution of the Backlund transformation is as follows
\begin {eqnarray}
q_{2n}=\frac{D_1^{n-1}D_1^n}{(D_0^n)^2} ,\qquad
r_{2n}=\frac{D_0^{n-1}D_0^n}{(D_1^n)^2},  \nonumber\\
q_{2n+1}=\frac{D_0^{n-1}D_0^n}{(D_0^n)^2}  ,\qquad
r_{2n+1}=\frac{D_1^{n-1}D_1^n}{(D_0^{n+1})^2} .
\end{eqnarray}
3. One- dimensional Heisenberg ferromagnetic in classical region (XXX - model).
$$
S'  =  [S,\ddot S], \qquad S=(S_-,S_0,S_+), \qquad  S_0^2+S_-S_+=1;
$$
$$
\tilde S_{-}  = S_{-}+2\left(\frac{1}{(\frac{s_+}{1+s_0})^\cdot}\right)^\cdot,
\quad
\tilde S_{+}  =  S_{+}+2\left(\frac{1}{(\frac{s_-}{1-s_0})^\cdot}\right)^\cdot,
$$
$$
\tilde S_0+1  =  -\tilde S_{-}\frac{S_{+}}{1+S_0}, \qquad
S^0_{-}  =0, \qquad  S^0_0=1, S_{+}'=2\ddot S_{+}
$$
\begin{eqnarray}
S_{-}^n &  = &  \frac{D_2^{n-1}D_2^n}{(D^n_1)^2}, \qquad
S_{0}^n+1=2\frac{D_0^{n}D_2^n}{(D^n_1)^2}, \nonumber\\
S_{0}^n & - &  1=2\frac{D_2^{n-1}D_0^{n+1}}{(D^n_1)^2}, \qquad
S_{+}^n=-4\frac{D_0^{n+1}D_0^n}{(D^n_1)^2}.
\end{eqnarray}

4.XYZ-model in classical region. The Landau-Lifshits equation.
$$
\vec S'=\vec S \times \ddot S+\vec S \times \vec (JS)
$$
$$
\vec S=(S_1,S_2,S_3), (\vec S)^2=1 , \quad  J=diag (J_1,J_2,J_3)
$$
  Under the steriographic projection
$$
u={S_1+iS_2\over 1+S_3} \qquad v={S_1-iS_2\over 1+S_3}
$$
and exchanging $ -it\to t $ it became a system of the equations:
$$
u'+\ddot u-2v{(\dot u)^2+R(u)\over1+uv}+{1\over2}\frac{\partial}{\partial u}R
(u)=0
$$
\begin{equation}
{}\label{10}
\end{equation}
$$
-v'+\ddot v-2u{(\dot v)^2+R(v)\over1+uv}+{1\over2}\frac{\partial}{\partial v}R
(v)=0
$$
where $ R(x)=\alpha x^4+\gamma x^2+\alpha \quad \frac{\partial R}{\partial x}=
4\alpha x^3+2\gamma x=2{R+\alpha(x^4-1)\over x}    \quad
\alpha={J_2-J_1\over 4}    \quad  \gamma={J_1+J_2\over2}-J_3 $
  The system (\ref{10}) is invariant under transformation $ u\to U,v\to V $:
\begin{equation}
U={1\over v} \qquad {1\over 1+{V\over v}}-{1\over 1+u v}=\frac
{v\ddot v-(\dot v)^2+\alpha(v^4-1)}{\dot v)^2+R(v)}\label{11}
\end{equation}
which is the Backlund transformation for this system.

5. Lund--Pohlmeyer- Regge model
\begin{eqnarray}
\dot y' - 4y + 2(xy)y' = & 0 &, \;
\tilde x = (\dot y + xy^2)^{-1} ,\nonumber\\
\dot x' - 4x - 2(xy)y' = & 0 &, \;
\tilde y = -\dot {(\dot y + xy^2)} + y^{-1}(\dot y +xy^2)^2 ;
\end{eqnarray}
\[
x_0 = 0  \;, \qquad \dot y'_0 = 4y_0.
\]
\begin{equation}
x_n=(-1)^{n+1}\frac{D^{n-1}_1}{D^n_1} \; ,\qquad
y_n=(-1)^n\frac{D^{n+1}_0}{D^n_0}.
\end{equation}
6. The main chiral field problem in a space of $n$ dimensions
(the case of an algebra $A_1$).
The main chiral field problem in $n$-dimensional space is described
by the following system of equations:
\begin{equation}
(\theta_i - \theta_j)\frac{\partial^2 f}{\partial x_i \partial x_j}=
[\frac{\partial f}{\partial x_i}, \frac{\partial f}{\partial x_j}],
\end{equation}
where the function f takes values in $A_1$ algebra,
$\theta_i$ are numerical parameters.
\begin{eqnarray}
F_{-} & = & \frac{1}{f_{+}} , \nonumber\\
\frac{\partial F_0}{\partial x_i} & = & (f_0 -F_0 +\theta_i)
\frac{\partial}{\partial x_i} \mbox{ln}f_+ -
\frac{\partial f_0}{\partial x_i}, \nonumber\\
\frac{\partial F_+}{\partial x_i} & = & (f_0 - F_0 +\theta_i)^2
\frac{\partial f_+}{\partial x_i} -2f_{+}(f_0 -F_0 +\theta_i)
\frac{\partial f_0}{\partial x_i} -f^2_+ \frac{\partial f_-}{\partial x_i}.
\end{eqnarray}.
The last equations can be rewritten in the matrix form
\begin{eqnarray}
\frac{\partial F}{\partial x_i} =  & \exp & [-X^+ (f_0-F_0+\theta_i)
f_+] \exp [H \mbox{ln}f_+] \; r
\frac{\partial f}{\partial x_i} \; r^{-1} \nonumber\\
& \exp & [-H \mbox{ln}f_+]
\exp [ X^+(f_0-F_0+\theta_i) f_+],
\end{eqnarray}
where $r$ is an automorphism of the algebra $ A_1 $ with the properties
\[
rX^{\pm}r^{-1}=-X^{\mp}, \; rHr^{-1}=-H;
\]
\[
f_{-}^0=0, \qquad f_0^0=\tau,  \qquad f^0_+=\alpha^0,
\]
where
\begin{equation}
\frac{\partial^2 \tau}{\partial x_i \partial x_j}=0, \;
(\theta_i - \theta_j)\frac{\partial^2 \alpha^0}{\partial x_i \partial x_j}=
2 [\frac{\partial \tau}{\partial x_i} \frac{\partial \alpha^0}{\partial x_j}-
\frac{\partial \tau}{\partial x_j} \frac{\partial \alpha^0}{\partial x_i}].
\end{equation}
To solve the Backlund transformation, let us consider the linear system
of equations:
\begin{equation}
\theta_i \frac{\partial\alpha^l}{\partial x_i}-
2 \frac{\partial \tau}{\partial x_i}\alpha^l=\frac{\partial\alpha^{l+1}}
{\partial x_i}.
\end{equation}
{}From the last equations it follows that each function $\alpha^l$ is
a solution of the equation for $\alpha^0$. We have an explicate
expression for $\alpha^s$:
\begin{equation}
\tau =\sum\phi_i(x_i), \;  \alpha^s=\int d \lambda (\lambda )^s
c(\lambda )
\exp (\sum \frac{\phi_i(x_i)}{\lambda -\theta_i}).
\end{equation}
In terms of the $\alpha^l$ the Backlund transformation have the solution
\begin{equation}
f^n_0=\frac{D^{n-1}_0}{D^n_0},\; f^n_0=\tau -\frac{\dot D_0^n}{D^n_0},\;
f^n_{+}=\frac{D^{n+1}_0}{D^n_0}\:.
\end{equation}
In the determinant $ \dot{D^n_0} $ numbering of the indices of the last row
is enlarged by unity.

7. The main chiral field problem for an arbitrary semisimple Lie algebra.

For the case of a semisimple Lie algebra and for an element $f$
being a solution of (12), the following statement takes place:
{\it There exists such an element $S$ taking values in a gauge group that}
\begin{equation}
S^{-1} \frac{\partial S}{\partial x_i} = \frac{1}{f_-}[\frac
{\partial f}{\partial x_i}, X^+_M] - \theta_i\frac{\partial}{\partial x_i}
\frac{1}{f_-}X^+_M.
\end{equation}
Here $X^+_M$ is the element of the algebra corresponding
to its maximal root divided by its norm, i.e.,
\[
[X^+_{M}, X^-] = H, \;
[H, X^{\pm}] = \pm 2 X^{\pm},
\]
$ -f_-$ is the coefficient
function in the decomposition of $f$ of the element corresponding to
the minimal root of the algebra.
In this terms the Backlund transformation is as follows:
\begin{equation}
\frac{\partial F}{\partial x_i} = S \frac{\partial f}{\partial x_i} S^{-1} +
\theta_i \frac{\partial S}{\partial x_i} S^{-1},
\end{equation}

8. The system of self-dual equations in four-dimensional space
(the case of the algebra $A_1$).
The self-dual equations for an element $f$ with values in a semisimple
Lie algebra  have the  following form:
\begin{equation}
\frac{\partial ^2 f}{\partial y \partial \bar { y}} +
\frac{\partial ^2 f}{\partial z \partial \bar { z}} =
[\frac{\partial f}{\partial y}, \frac{\partial f}{\partial z}].\label{23}
\end{equation}
For this system the Backlund transformation has the form
\[
F_{-} =\frac{1}{f_{-}}
\]
\begin{eqnarray}
\frac{\partial}{\partial y}F_0  & = & \frac{\partial}{\partial \bar {z} }
\mbox{ln} f_-  -
\frac{\partial}{\partial y} f_0  + ( f_0 - F_0 )
\frac{\partial}{\partial y} \mbox{ln}f_-,\nonumber\\
\frac{\partial}{\partial z}F_0  & = -& \frac{\partial}{\partial \bar {y} }
\mbox{ln} f_-  -
\frac{\partial}{\partial z} f_0  + ( f_0 - F_0 )
\frac{\partial}{\partial z} \mbox{ln}f_-,\nonumber\\
\frac{\partial}{\partial y}F_+ &  =  & -f_-\{ (f_0-F_0)
\frac{\partial}{\partial y}
(f_0 - F_0) + \frac{\partial}{\partial \bar {z}}
(f_0 - F_0) \} - f^2_- \frac{\partial}{\partial y}f_+ ,\nonumber\\
\frac{\partial}{\partial z}F_+ &  =  & -f_-\{ (f_0-F_0)
\frac{\partial}{\partial z}
(f_0 - F_0) - \frac{\partial}{\partial \bar {y}}
(f_0 - F_0) \} - f^2_- \frac{\partial}{\partial z}f_+.\label{24}
\end{eqnarray}
The substitution  of (\ref{24}) in the density of the topological charge
gives the equality:
\[
Q_F = q_f + \framebox(10,10){}\; \framebox(10,10){}
\; \mbox{ln} f_-
\]
For the integration of the Backlund transformation we have the linear
system of equation
\begin{equation}
\frac{\partial\alpha^l}{\partial \bar{y}}+
2 \frac{\partial \tau}{\partial z}\alpha^l=\frac{\partial\alpha^{l+1}}
{\partial z} \;,
\frac{\partial\alpha^l}{\partial\bar{z}}-
2 \frac{\partial \tau}{\partial y}\alpha =\frac{\partial\alpha^{l+1}}
{\partial y}\label{25}
\end{equation}
In this terms we have the solution of the self-dual system
\begin{equation}
f^n_{-}=\frac{D_0^{n-1}}{D_0^n},\; f_0^n=\frac{\dot D_0^n}
{D_0^n}+\tau ,\; f_+^0=\frac{D_0^{n+1}}{D_0^n}.\label{26}
\end{equation}

9. The system of self-dual equations in the case of an arbitrary semisimple
algebra.

The following statement takes place:

{\it There exists such an element $S$ taking the values in the gauge group,
that}
\begin{eqnarray}
S^{-1} \frac{\partial S}{\partial y} & = & \frac{1}{f_-}[\frac
{\partial f}{\partial y}, X^+_M] - \frac{\partial}{\partial\bar z}
(\frac{1}{f_-}) X^+_M, \nonumber\\
S^{-1} \frac{\partial S}{\partial z} & = & \frac{1}{f_-}[\frac
{\partial f}{\partial z}, X^+_M] + \frac{\partial}{\partial\bar y}
(\frac{1}{f_-}) X^+_M. \label{27}
\end{eqnarray}
Here $X^+_M$ is the element of the algebra corresponding to its
maximal root, divided by its norm, i.e.,
\[
[X^+_{M}, X^-] = H, \;
[H, X^{\pm}] = \pm 2 X^{\pm} ,
\]
{}$-f_{-}$ is the coefficient
function in the decomposition of $f$ on the element corresponding to
the minimal root of the algebra.
The Backlund transformation have the form
\begin{equation}
\frac{\partial F}{\partial y} = S \frac{\partial f}{\partial y} S^{-1} +
\frac{\partial S}{\partial \bar z} S^{-1},\qquad
\frac{\partial F}{\partial z} = S \frac{\partial f}{\partial z} S^{-1} -
\frac{\partial S}{\partial \bar y} S^{-1}.\label{28}
\end{equation}
10. The main chiral field problem with the moving poles.
Many integrable systems arise from (\ref{23}) by imposing
symmetry requirements on the solution. The cylindrically
symmetric condition in four dimensional space restricts
the form of the function $f$,
\[
f = \frac{1}{\bar y}f(\xi,\bar \xi), \qquad \xi =
\frac{z- \bar z}{2} + [(\frac
{z + \bar z}{2})^2 + y \bar y]^{1/2},  \qquad \bar\xi=-\xi^*.
\]
\begin{equation}
(\xi - \bar \xi )\frac{\partial^2}{\partial \xi \partial\bar \xi}f=
[\frac{\partial}{\partial \xi}f, \frac{\partial}{\partial \bar \xi}f].
\end{equation}
This is the equation for the main chiral field with moving poles.

The result of integration of equation (\ref{27}) is given in the form
\[
S=S(\xi,\bar \xi ) \exp X^+_M \frac{z}{f_-};
\]
\begin{eqnarray}
S^{-1}\frac{\partial}{\partial \xi}S  & = &
\frac{1}{f_-}[\frac{\partial}
{\partial \xi}f,  X^+_M] - \xi \frac{\partial}{\partial \xi}
\frac{1}{f_-}X^+_M, \nonumber\\
S^{-1}\frac{\partial}{\partial \bar \xi}S  & = &
\frac{1}{f_-}[\frac{\partial}
{\partial \bar \xi}f,  X^+_M] - \bar \xi \frac{\partial}{\partial
\bar \xi}
\frac{1}{f_-}X^+_M,
\label{30}
\end{eqnarray}
and the Backlund transformation has the following form:
\begin{equation}
\frac{\partial}{\partial \xi}F = S (\frac{\partial}{\partial \xi}f) S^{-1} -
\xi \frac{\partial S}{\partial \xi} S^{-1}, \quad
\frac{\partial}{\partial\bar \xi}F = S(\frac{\partial}{\partial
\bar \xi}f) S^{-1} -
\bar \xi \frac{\partial S}{\partial \bar \xi} S^{-1}.
\label{31}
\end{equation}
The relations (\ref{30}), (\ref{31}) realize the Backlund transformation
for the main chiral field with moving poles.

11. The self dual equation under of cylindrical symmetric in
three dimensional space
  The condition of cylindrical symmetric in three dimensional space
leads to the following form of the solution to equation :
\[
f= \frac{1}{\bar y} f(\xi, \bar \xi), \qquad  \xi=
\frac{z+ \bar z}{2} + i (y \bar y)^{1/2}, \qquad
\bar \xi = - \xi^*;
\]
\begin{equation}
(\xi - \bar \xi) \frac{\partial^2 f}{\partial \xi \partial \bar \xi} =
\frac{1}{2} (\frac{\partial f}{\partial  \bar \xi} -
\frac{\partial f}{\partial  \xi}) + [\frac{\partial f}{\partial  \bar \xi}
\frac{\partial f}{\partial  \xi}].
\label{32}
\end{equation}

The Backlund transformation for equation (\ref{32}) arising from
(\ref{27}), (\ref{28}) has the form
\begin{eqnarray}
S^{-1}\frac{\partial}{\partial \xi}S  & = &
\frac{1}{f_-}[\frac{\partial}
{\partial \xi}f,  X^+_M] + (\frac{1}{f_-}-
\frac{\xi -\bar \xi}{2} \;\frac{\partial}{\partial \xi}
\frac{1}{f_-}) X^+_M, \nonumber\\
S^{-1}\frac{\partial}{\partial \bar \xi}S  & = &
\frac{1}{f_-}[\frac{\partial}
{\partial \bar \xi}f,  X^+_M] + (\frac{1}{f_-}-
 \frac{\xi -\bar \xi}{2} \;\frac{\partial}{\partial
\bar \xi}
\frac{1}{f_-}) X^+_M, \nonumber
\end{eqnarray}
\[
\frac{\partial}{\partial \xi}F = S (\frac{\partial}{\partial \xi}f) S^{-1} +
\frac{\bar \xi-\xi}{2} \; \frac{\partial S}{\partial \xi} S^{-1}, \quad
\frac{\partial}{\partial\bar \xi}F = S(\frac{\partial}{\partial
\bar \xi}f) S^{-1} -
\frac{\bar \xi- \xi}{2} \;\frac{\partial S}{\partial \bar \xi} S^{-1}.
\]

In the case of algebra $A_1$ system (\ref{32}) arises in the
integration problem of the general relativity  with two commuting
Killing vectors.

12. The cylindrical symmetrically solution invariant under two orthogonal
four-dimensional axis.
\begin{equation}
x_1\frac{\partial^2F}{\partial x_1^2}+x_2\frac{\partial^2F}{\partial x_2^2}=
[\frac{\partial F}{\partial x_1},\frac{\partial F}{\partial x_2}]\label{10.80}
\end{equation}
  The case of the algebra $ A_1 $ the explicate form of the Backlund
transformation is the following $ F_{-}={1\over f_{-}} $
$$
\frac{\partial F_0}{\partial x_1}=-1+(f_0-F_0)\frac{\partial lnf_{-}}
{\partial x_1}+x_2\frac{\partial lnf_{-}}{\partial x_2}-\frac{\partial f_0}
{\partial x_1}
$$
$$
\frac{\partial  F_0}{\partial x_2}=1+(f_0-F_0)\frac{\partial lnf_{-}}
{\partial x_2}-x_1\frac{\partial lnf_{-}}{\partial x_1}-\frac{\partial f_0}
{\partial x_2}
$$
\begin{equation}
{}
\end{equation}
$$
\frac{\partial F_{+}}{\partial x_1}=-f_{-}[(f_0-F_0)\frac{\partial(f_0-F_0)}
{\partial x_1}+x_2\frac{\partial(f_0-F_0)}{\partial x_2}]-f_{-}^2
\frac{\partial f_{+}}{\partial x_1}
$$
$$
\frac{\partial F_{+}}{\partial x_2}=-f_{-}[(f_0-F_0)\frac{\partial (f_0-F_0)}
{\partial x_2}-x_1\frac{\partial (f_0-F_0)}{\partial x_1}]-f_{-}^2
\frac{\partial f_{+}}{\partial x_2}
$$
  The linear system of the equations has the form
$$
x_1\frac{\partial \alpha^l}{\partial x_1}-\alpha^l+2\alpha^l
\frac{\partial \tau^l}{\partial x_2}=\frac{\partial \alpha^{l+1}}
{\partial x_2}
$$
\begin{equation}
x_2\frac{\partial \alpha^l}{\partial x_2}-\alpha^l-2\alpha^l
\frac{\partial \tau^l}{\partial x_1}=-\frac{\partial \alpha^{l+1}}
{\partial x_1}
\end{equation}
$$
\tau^{l+1}=\tau^l+{x_1-x_2 \over 2}
$$
The solution of the Backlund transformation coincide with the self-dual case
(see (\ref{25}) and (\ref{26})).

4.Let denote by $ \theta_i $ the solution of the integrable system, which arise
after application i-times the Backlund transformation to some given one.
The Backlund transformation connect the solutions with different values of
the index i and so it arise the infinite chain of equations. In some cases
this chain may be limited from "left" or from the "right" and as we have seen
in the pervious section this possibility lead to some subclass of the solution
of the integrable systems, which depend from the definite number of the
arbitrary functions. The other possibility consists in the assuming of the
periodical condition of the solution $ \theta_{N+i}=\theta_i $, which choice
some finite subclass of the solutions of the initial infinite chain. Let
consider this possibility on the examples of the nonlinear Schrodinger
equations (see (\ref{4})-(\ref{7})).

In that case we have only one variable for every value of the index i $ r_i
\equiv \exp x_i $. In this notations the Backlund transformations are rewritten
as follows

a)
\[
q_{i+1}={1\over r_i}=\exp-x_i,\quad r_{i+1}=r_i[q_ir_i-\ddot{\mbox{ln} r_i}]
\]
or
\[
\ddot x_i+\exp (x_{i+1}-x_i)-\exp (x_i-x_{i-1})=0
\]
Under the condition $ x_{i+N}=x_i $ this is the equations of the periodical
Toda lattice for the algebra $ A_N $

b)

\[
(\ln \dot x_i)\dot {} + \exp (x_{i+1}-x_i) + \exp (x_i-x_{i-1})=0
\]

c)
In that case $ q_{i+1}=\exp x_i $ and for the variable $ N_i\equiv r_i $ we
receive the chain of the equations
\[
\dot N_i=N_i^2(N_{i-1}-N_{i+1})
\]
This is the Lotky-Volterra chain.
In the work \cite{4} the independent investigation of the periodical
integrable chains was used for the construction of explicate form of the
Backlund transformation for the definite class of the integrable systems.
The solution of the Backlund transformation under the periodical condition may
be connected with the double periodical solutions of such systems.

5. The main result of the present paper is in the formulas
of the section 3 which realized the Backlund transformation for
the integrable system under consideration. A more important
consequences is in the assumption of a possibility
to obtain the Backlund transformation on the base of the group theory.
The Backlund transformation in question is the mapping
whose invariants are the equations for integrable system.
For this reason an independent construction of the Backlund transformation
is equivalent to the problem of enumeration
of all integrable systems or some class of them.


\begin{thebibliography}{**}
\bibitem{1}
S. M. Chumakov, A. N. Leznov, and V. I. Man'ko. Trudi PHIAN {\bf 167}
(1986), 232-237.
\bibitem{2}
A.N.Leznov and M.V.Saveliev Acta. Appl. Math.v.16.(1-74) 1991.
\bibitem{3}
A. N. Leznov. Preprint IHEP 91-71 1991.
\bibitem{4}
A.B.Shabat and R.I.Yamilov Leningrad Math.J. Vol.2(1991),No.2.

\end{thebibliography}
\end{document}